\documentclass[graybox]{svmult}

\usepackage{mathptmx}       
\usepackage{helvet}         
\usepackage{courier}        
\usepackage{type1cm}        
\usepackage{makeidx}         
\usepackage{graphicx}        
\usepackage{multicol}        
\usepackage[bottom]{footmisc}

\makeindex             

\begin{document}

\title*{The energy cascade of surface wave turbulence: toward identifying the active wave coupling}
 \titlerunning{The energy cascade of surface wave turbulence}
\author{Antoine Campagne, Roumaissa Hassaini, Ivan Redor, Joel Sommeria and Nicolas Mordant}
\institute{Antoine Campagne \at LEGI, Grenoble, France, \email{antoine.campagne@univ-grenoble-alpes.fr}
\and Roumaissa Hassaini \at LEGI, Grenoble, France, \email{roumaissa.hassaini@univ-grenoble-alpes.fr}
\and Ivan Redor \at LEGI, Grenoble, France, \email{ivan.redor@univ-grenoble-alpes.fr}
\and Joel Sommeria \at LEGI, Grenoble, France, \email{joel.sommeria@legi.cnrs.fr}
\and Nicolas Mordant (picture) \at LEGI, Grenoble, France, \email{nicolas.mordant@univ-grenoble-alpes.fr}}
%
%
\maketitle

\abstract{We investigate experimentally turbulence of surface gravity waves in the Coriolis facility in Grenoble by using both high sensitivity local probes and a time and space resolved stereoscopic reconstruction of the water surface. We show that the water deformation is made of the superposition of weakly nonlinear waves following the linear dispersion relation and of bound waves resulting from non resonant triadic interaction. Although the theory predicts a 4-wave resonant coupling supporting the presence of an inverse cascade of wave action, we do not observe such inverse cascade. We investigate 4-wave coupling by computing the tricoherence i.e. 4-wave correlations. We observed very weak values of the tricoherence at the frequencies excited on the linear dispersion relation that are consistent with the hypothesis of weak coupling underlying the weak turbulence theory.}

\section{Introduction}

Wave Turbulence is a general framework that aims at describing the statistical properties of a large ensemble of waves. Although no general theory exists, the Weak Turbulence Theory (WTT) focusses on the case of vanishing non linearity in very large systems \cite{Zakh,Naz,New}. It predicts an energy cascade in scale space between the large scale of forcing down to small scales at which dissipation dominates. Due to weak nonlinearity energy transfer occurs among resonant waves. Oceanic waves is the natural field of application of the theory following the work of Hasselman \cite{Has} that assumes transfer among 4 resonant waves. A major result of the WTT is that analytic solutions of the wave Fourier spectrum can be exhibited in many cases, the so-called Kolmogorov-Zakharov spectra. For gravity surface waves the prediction of the wave elevation spectrum is \cite{Naz}:
\begin{eqnarray}
E^{\eta}(k)&\propto& g^{1/2}P^{1/3}k^{-5/2}\\
\textrm{or }E^{\eta}(\omega)&\propto& gP^{1/3}\omega^{-4}
\end{eqnarray}
where $g$ is the gravity acceleration, $P$ is the energy flux, $k$ is the wave number and $\omega$ the angular frequency. Although some field measurements of the spectra appear compatible with this prediction, laboratory experiments fail to reproduce this prediction. The observed spectral exponents of the frequency spectrum are significantly steeper than the $-4$ theoretical value \cite{NaLu,Deik,Aub2}. Our goal is to investigate further the statistical properties of the wave field recorded experimentally to obtain some insight on the reasons for the discrepancy between theory, observations and laboratory data. For surface gravity wave, due to the 4-wave coupling, the theory predicts also an inverse cascade of wave action \cite{Naz} that maybe responsible for the long wave generation by the wind.

\section{Experimental setup}

\begin{figure}[!htb]
\sidecaption
\includegraphics[width=6cm]{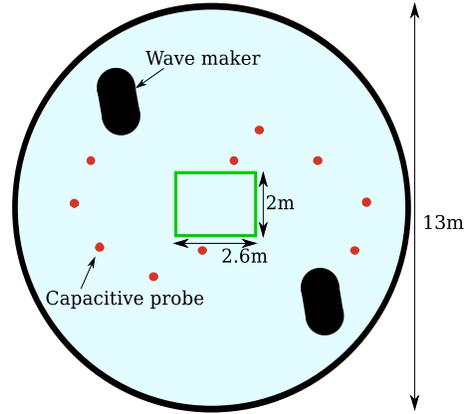}
\caption{Schematics of the setup in the Coriolis facility. The tank is 13m in diameter and the water is 0.9~m deep. The position of the two wavemakers is shown as black ovals and that of the 10 capacitive probes is shown as red dots. The field of view of the stereoscopic reconstruction is the green rectangle at the center.}
\label{setup}
\end{figure}
Waves are generated by two wedge wavemakers in a circular tank of 13~m diameter and 0.9~m depth (the Coriolis facility located in Grenoble, France). Wave elevation is recorded by a set of 10 capacitive wave gauges that provide a local measurement and a stereoscopic system that provides a space and time resolved measurement of the wave elevation over a surface $2\times 2.6$~m$^2$ (Fig.~\ref{setup}). 

\begin{table}
\caption{Parameters of the three datasets. $f_p$ is the frequency of the peak of the spectrum. $k_p$ is the wavenumber corresponding to $f_p$ following the linear dispersion relation. $\sigma_\eta$ is the elevation variance. $\epsilon_p$ is the wave steepness computed as $\epsilon_p=2k_p\sigma_\eta$ (see \cite{Aub2}).}
\label{tab}       
%
%
\begin{tabular}{p{2cm}p{2.4cm}p{2cm}p{2cm}p{2cm}}
\hline\noalign{\smallskip}
Dataset & $f_p$ [Hz] & $k_p$ [m$^{-1}$] & $\sigma_\eta$ [m] & $\epsilon_p$  \\
\noalign{\smallskip}\svhline\noalign{\smallskip}
     weak & 0.65 & 1.83 & 0.0294 & 0.11\\
     strong & 0.76 & 2.4 & 0.0339 & 0.16\\   
     short & 1.5 & 9.05 & 0.0131 & 0.24  \\       
\noalign{\smallskip}\hline\noalign{\smallskip}
\end{tabular}
\end{table}

Three datasets are acquired with distinct generation and called \emph{weak}, \emph{strong} and \emph{short} (see Table \ref{tab}). The waves are generated by oscillating the wavemakers with a random modulation of amplitude $\pm 0.15$ Hz around a central frequency $f_p$ with a vertical amplitude of 2~cm. The \emph{weak} dataset corresponds to the lowest peak frequency and a moderate wave steepness $\epsilon_p=0.11$. In the \emph{strong} dataset the wavemakers are operated at a slightly higher frequency at which they are more efficient and thus the steepness is larger $\epsilon_p=0.16$. The \emph{short} datasets corresponds to smaller wavemakers that are operated at a higher frequency ($1.5$~Hz) so that to generate shorter waves and investigate the possible generation of an inverse cascade. The stepness is very large in this case.

\section{Fourier spectra}

\begin{figure}[!htb]
\sidecaption
\includegraphics[width=7.5cm]{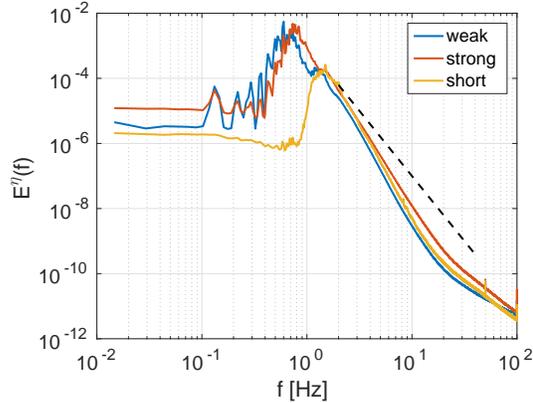}
\caption{Frequency spectrum $E^\eta(f)$ of the surface elevation for the three datasets.}
\label{sp}
\end{figure}

Frequency Fourier spectra are shown in Fig.~\ref{sp} for all three datasets. A turbulent spectrum is generated at frequencies higher that the peak frequency. At change of slope is observed at 14~Hz that corresponds to the gravity-capillary crossover. At frequencies lower than the peak frequency, the wave spectrum is two orders of magnitude lower than at the peak frequency meaning that no inverse cascade is observed even for the \emph{short} dataset. This observation raises the question of the relevance of the 4-wave coupling in our experiment.

\begin{figure}[!htb]
\sidecaption
\includegraphics[width=7.5cm]{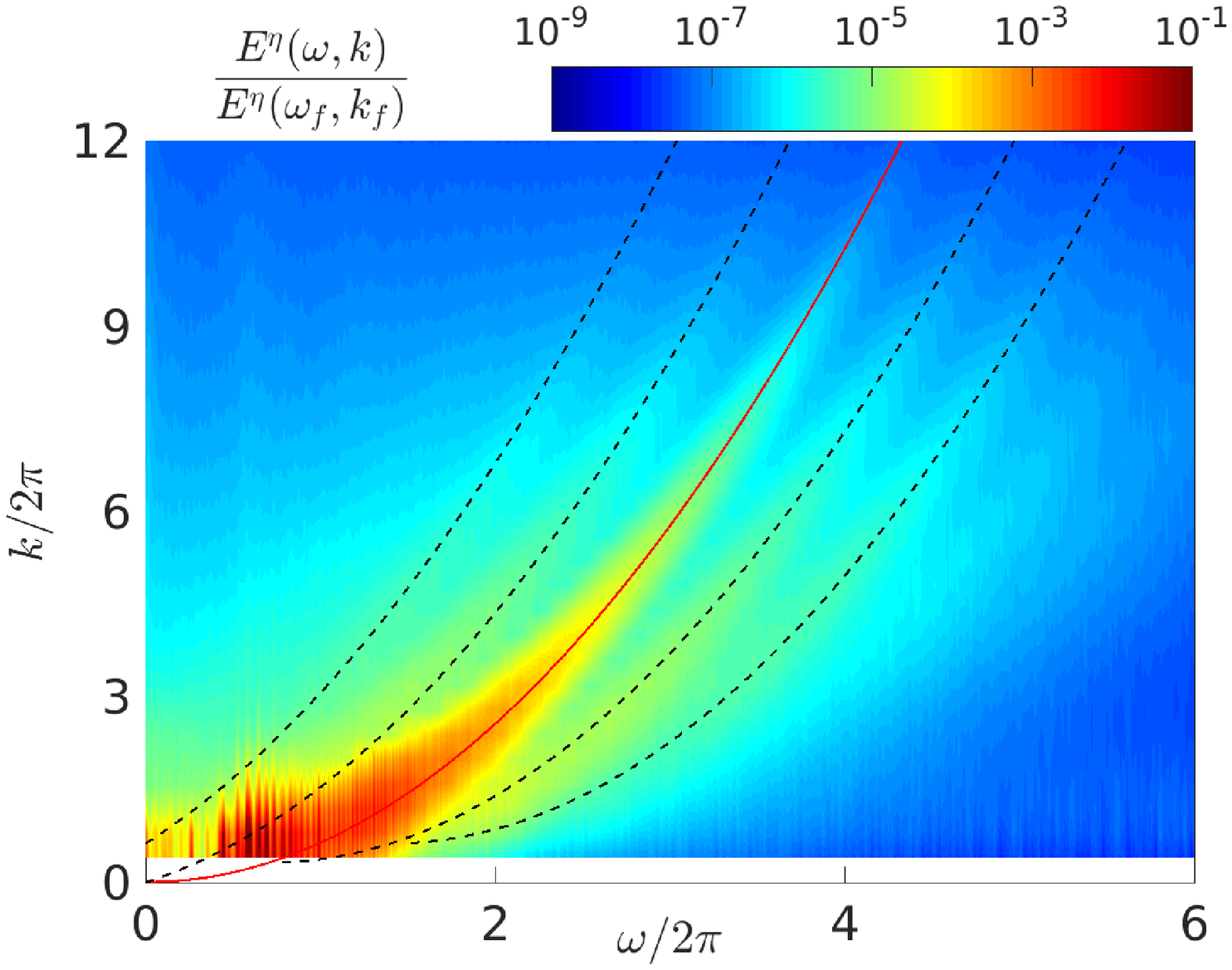}
\caption{$k-\omega$ spectrum $E^\eta(\omega,k)$ for $\epsilon=0.11$ (``weak'' case). The spectrum has been normalized by its value at the forcing and is displayed in $\log_{10}$ scale. The red line is the linear dispersion relation (\ref{ldr}) and the dashed lines corresponds to bound waves (\ref{bound}) (see text).}
\label{kw}
\end{figure}
\begin{figure}[!htb]
\sidecaption
\includegraphics[width=7.5cm]{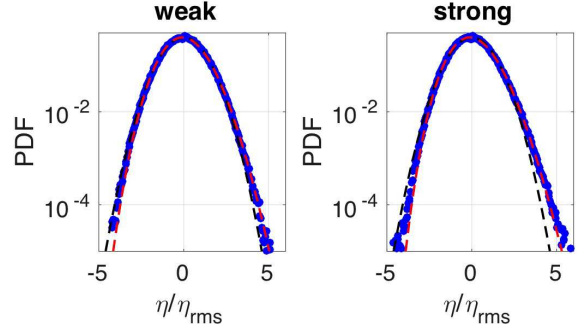}
\caption{Distribution of the wave elevation for the \emph{weak} (left) and \emph{strong} (right) case. The black dashed line is a Gaussian distribution and the red dashed line is the Tayfun distribution that corresponds to the parameter of each case.}
\label{pdf}      
\end{figure}

Figure~\ref{kw} displays the full frequency-wavenumber spectrum $E^\eta(\omega,k)$ corresponding to the \emph{weak} case. A strong concentration of the energy is observed on the linear dispersion relation (LDR, red curve) that follows:
\begin{equation}
\omega_{LDR}=\sqrt{gk+\frac{\gamma k^3}{\rho}}\, ,
\label{ldr}
\end{equation}
where  $\omega_{LDR}$ is the angular frequency, $k$ is the wavenumber, $g$ is the gravity acceleration, $\gamma$ is the surface tension and $\rho$ is the density of water. This shows that most of the energy is made of freely propagating waves. Nevertheless a significant amount of the energy lies out of the linear dispersion relation. In particular several lines can be distinguished that are highlighted by the black dashed lines. These lines where constructed by translating the dispersion relation by multiples of the forcing peak so that the equations of the dashed lines are
\begin{equation}
\omega^{(\pm n)}=\omega_{LDR}(k\mp nk_p)\pm n\omega_p\, ,
\label{bound}
\end{equation}
with $n$ being an integer either positive or negative. For instance for $n=\pm1$, energy can be transferred on the first dashed line (on the left or right of the LDR) by triadic interactions between a free wave that lies on the peak of the spectrum at position $(\omega_p, k_p)$ and another wave on the LDR $(\omega_{LDR}(k),k)$. Energy is thus transferred to position $(\omega_{LDR}(k)\pm\omega_p,k\pm k_p)$. For $|n|>1$, the same process implies successive harmonics of the forcing peak. These waves are not free to propagate and are known as bound waves. A first effect of the bound waves is that the statistics of the wave elevation are not Gaussian. Indeed as seen in Fig.~\ref{pdf}, the distribution of the wave elevation is positively skewed and follows the Tayfun distribution which is known to incorporate second order effets \cite{Soc}. The asymmetry is more pronounced for the \emph{strong} dataset for which the slope is higher than for the \emph{weak} case.

\section{Occurrence of 4-wave correlations}

\begin{figure}[!htb]
\includegraphics[width=\textwidth]{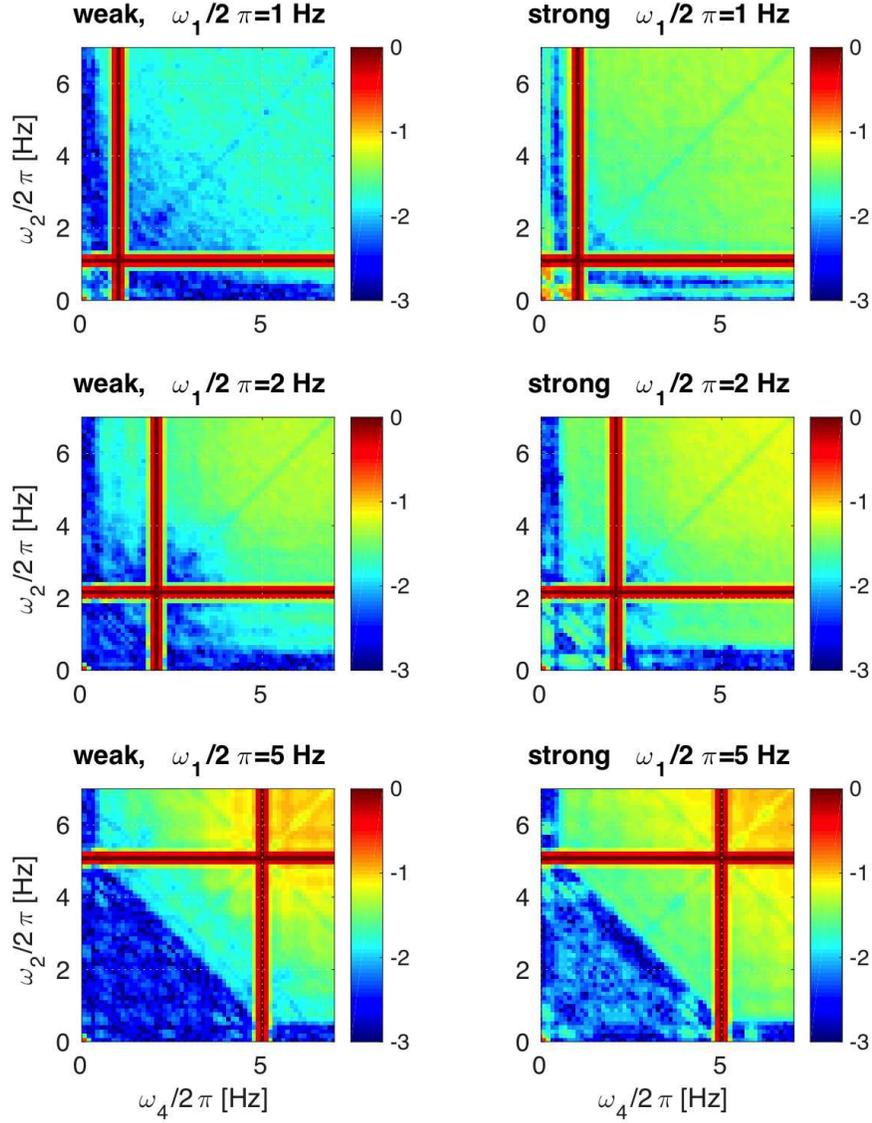}
\caption{Maps of the tricoherence computed on the local probe data following (\ref{tri}) for 3 given values $\omega_1/2\pi=1$, 2 and 5~Hz from top to bottom. The left column corresponds to the \emph{weak} dataset and the right column to the \emph{strong} one. Colors correspond to $\log_{10} C$. See text for details.}
\label{trico}       
\end{figure}

In the WTT for gravity waves, 3-wave coupling is not resonant as the bound wave is not a true free wave and thus it is not expected to contribute directly to the energy cascade~\cite{Has}. Indeed for deep water the resonance equations
\begin{equation}
\omega_{LDR1}+\omega_{LDR2}=\omega_{LDR3}\quad \textrm{,} \quad \mathbf k_1+\mathbf k_2=\mathbf k_3
\end{equation}
do not have non trivial solutions in the gravity range due to the curvature of the LDR. Thus the resonant transfers of the WTT occur through 4-wave resonant coupling.
\begin{equation}
\omega_{LDR1}+\omega_{LDR2}=\omega_{LDR3}+\omega_{LDR4}\quad \textrm{,} \quad \mathbf k_1+\mathbf k_2=\mathbf k_3+\mathbf k_4
\end{equation}
In order to fully express this 4-wave coupling in the theory, the 3-wave non resonant contributions are summed through a canonical change of variables \cite{Kra}. In this framework, part of the 4-wave coupling is actually due to the interplay of two non resonant triads. Unfortunately the change of variable is quite involved and is very hard to implement on experimental data. Nevertheless one can probe the occurrence of the 4-wave coupling through calculation of 4-wave tricoherence defined as
\begin{eqnarray}
&C(\omega_1,\omega_2,\omega_3,\omega_4)&=\frac{\langle\tilde\eta(\omega_1)\tilde\eta^\star(\omega_2)\tilde\eta(\omega_3)\tilde\eta^\star(\omega_4)\rangle}{\sqrt{\langle|\tilde\eta(\omega_1)\tilde\eta(\omega_3)|^2\rangle\langle|\tilde\eta(\omega_2)\tilde\eta(\omega_4)|^2\rangle}}\\
&\textrm{with }& \omega_{1}+\omega_{3}=\omega_{2}+\omega_{4} \nonumber \, .
\label{tri}
\end{eqnarray}
$\tilde\eta(\omega)$ is the Fourier transform in time of the elevation field at a given point over a temporal window of finite duration (chosen to 8.5~s). $\langle\,\,\rangle$ is an average over successive temporal windows and over the local probes. The four frequencies are imposed to be resonant so that the tricoherence $C(\omega_1,\omega_2,\omega_3,\omega_4)$ is actually depending only on three of the frequencies. The denominator is chosen to impose $|C|\leq1$ and that $C$ is nondimensional. Examples of the values of the tricoherence are shown in Fig.~\ref{trico} for the case \emph{weak} and \emph{strong}. The tricoherence being a 3D object we impose given values of $\omega_1$ (here $\omega_1/2\pi$ is 1, 2 or 5~Hz) chosen in the gravity range of frequencies. The statistical convergence level is estimated to be $3\,10^{-3}$ which corresponds to dark blue colors. The red cross corresponding to values of tricoherence equal to one are due to trivial combinations of the frequencies such as $\omega_1=\omega_2$ and $\omega_3=\omega_4$. Out of this special cases converged values of tricoherence can be observed. 

\begin{figure}[!htb]
\sidecaption
\includegraphics[width=7.5cm]{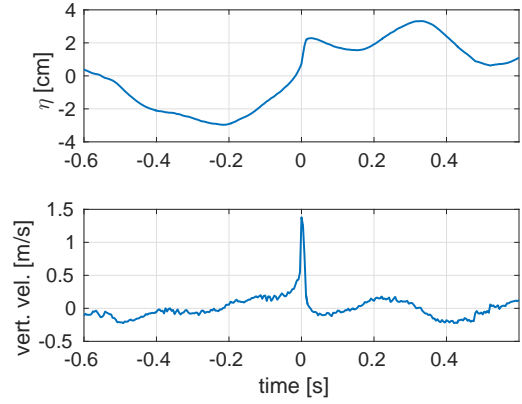}
\caption{Example of a singularity in the elevation due to a whitecapping at small scale. Top is the elevation and bottom is the vertical velocity.}
\label{def}      
\end{figure}
Note that small scale singular events can be also observed (see example in Fig.~\ref{def}). These events can be related to small whitecapping events of the waves due to the fact that nonlinearity is not vanishingly small. These events are quite rare and occur at relatively small scale as compared to the large scales associated to gravity waves that are discussed in the following. Thus, we detect them by thresholding the vertical velocity and do not take them into account in the computation of the tricoherence.

For $\omega_1/2\pi=1$~Hz, the coherence is very weak (about $10^{-2}$ for the \emph{weak} case) and seems to be almost zero (at our level of convergence) when either $\omega_2$ or $\omega_4$ is less than $\omega_1$. This appears consistent with the lack of observation of an inverse cascade. The coherence is increasing with $\omega_2$ and $\omega_4$. For $\omega_1/2\pi=5$~Hz it can even reach strong values close to $10^{-1}$ when $\omega_2/2\pi\approx\omega_4/2\pi\approx 7$~Hz. The frequencies get close to the gravity-capillary crossover (14~Hz) at which a specific 3-wave resonant process has been observed that involves one gravity wave and two capillary waves \cite{Aub1}. Such very strong values of the tricoherence may be related to this process rather than to resonance between four gravity waves. For $\omega_1/2\pi=1$~Hz and $1~\textrm{Hz}<\omega_2/2\pi,\omega_4/2\pi< 4$~Hz the observed non zero coherence maybe a trace of genuine 4-wave coupling that may be responsible for the direct transfert of energy along the dispersion relation as observed in Fig.~\ref{kw} at frequencies up to 4~Hz.

\section{Concluding remarks}
The analysis of the tricoherence suggests that a 4-wave resonant process maybe indeed operating at low frequencies between the forcing peak and 4~Hz and be responsible for the energy flux that provides energy along the dispersion relation as observed on the $(k,\omega)$ spectrum. At the lowest frequency the lack of coherence is consistent with the lack of inverse cascade. At the highest frequencies the very large values of the tricoherence are most likely due to a distinct 3-wave resonant process reported previousy by Aubourg \& Mordant \cite{Aub1} near the gravity-capillary crossover. As this process should be always operating, the condition for a clearer evidence of the 4-wave resonant process among gravity waves would require the use of a much larger wave facility.

\begin{acknowledgement}
This project has received funding from the European Research Council (ERC) under the European Union's Horizon 2020 research and innovation programme (grant agreement No 647018-WATU).\end{acknowledgement}
\end{document}